\documentclass[letterpaper]{article}
\usepackage[draft]{aaai_style/aaai25} 
\usepackage{times} 
\usepackage{helvet}
\usepackage{courier}
\usepackage[hyphens]{url}
\usepackage{graphicx}
\usepackage{natbib}
\usepackage{caption}
\usepackage{microtype}
\usepackage{graphicx}
\usepackage{subfigure}
\usepackage{booktabs, tabularx} 
\usepackage{multirow}
\newcolumntype{L}{>{\raggedright\arraybackslash}X}
\usepackage{enumitem}
\usepackage{amsmath}
\usepackage{amssymb}
\usepackage{mathtools}
\usepackage{amsthm}

\usepackage[capitalize,noabbrev]{cleveref}

\theoremstyle{plain}
\newtheorem{theorem}{Theorem}[section]

\theoremstyle{definition}
\newtheorem{definition}[theorem]{Definition}

\theoremstyle{remark}

\usepackage{layout}
\usepackage[dvipsnames]{xcolor}

\begin{document}

\title{Toward Valid Measurement Of (Un)fairness For Generative AI: \\A Proposal For Systematization Through The Lens Of Fair Equality of Chances}
\author {
    Kimberly Truong\textsuperscript{\rm 1},
    Annette Zimmermann\textsuperscript{\rm 2},
    Hoda Heidari\textsuperscript{\rm 1}
}
\affiliations {
    \textsuperscript{\rm 1}Carnegie Mellon University, \textsuperscript{\rm 2}University of Wisconsin-Madison\\
    kltruong@cmu.edu
}

\maketitle


\begin{abstract}

Disparities in the societal harms and impacts of Generative AI (GenAI) systems highlight the critical need for effective unfairness measurement approaches. While numerous benchmarks exist, designing \emph{valid} measurements requires proper systematization of the unfairness construct. 
Yet this process is often neglected, resulting in metrics that may mischaracterize unfairness by overlooking contextual nuances, thereby compromising the validity of the resulting measurements. Building on established (un)fairness measurement frameworks for predictive AI, this paper focuses on assessing and improving the \textit{validity} of the measurement task. By extending existing conceptual work in political philosophy, we propose a novel framework for evaluating GenAI unfairness measurement through the lens of the Fair Equality of Chances framework. Our framework decomposes unfairness into three core constituents: the \emph{harm/benefit} resulting from the system outcomes, \emph{morally arbitrary factors} that should not lead to inequality in the distribution of harm/benefit, and the \emph{morally decisive factors}, which distinguish subsets that can justifiably receive different treatments. By examining fairness through this structured lens, we integrate diverse notions of (un)fairness while accounting for the contextual dynamics that shape GenAI outcomes. We analyze factors contributing to each component and the appropriate processes to systematize and measure each in turn. This work establishes a foundation for developing more valid (un)fairness measurements for GenAI systems.
\end{abstract}

\section{Introduction}
\label{section:intro}

Generative AI (GenAI) systems have demonstrated concerning patterns of stereotypical, derogatory, exclusionary, and overall harmful outputs that disproportionately affect underprivileged and marginalized communities \citep{gallegos2024bias}. 
These incidents highlight the importance of unfairness and bias measurements for GenAI so that concerning disparities can be assessed and mitigated.
Recent literature has revealed (un)fairness measurements in theory and practice are misaligned \cite{harvey2024gaps}, with general-purpose benchmarks often failing to capture the contextual manifestations of unfairness in generative outputs, thereby misrepresenting the intended (un)fairness concept. 

\paragraph{The evaluation crisis in GenAI} 
Beyond measures of unfairness and bias, the literature on designing evaluation metrics and measures for GenAI systems is in its infancy. The sudden emergence of GenAI has transformed the technological landscape, creating a race among researchers and practitioners alike to keep up with rapid innovations in model development and data release \citep{bommasani2022opportunities}. This rapid progression has relegated proper measurement design to an afterthought, with many GenAI benchmarks released prematurely with minimal methodological backing or documentation. This oversight has made it increasingly challenging to select appropriate evaluation metrics--or determine if any are suitable--among the slew of existing benchmarks. 

\paragraph{Validity issues with existing metrics}
Well-established theories of measurement from the social sciences emphasize that \emph{validity} should be a core consideration in the design of any measurement \citep{drost2011validity}, yet many existing GenAI evaluation measurements suffer from validity issues; that is, they fail to accurately capture the concept they intend to study \cite{coston2023validity}. Valid unfairness metrics must be context-specific \cite{al2024ethical}, yet static, general-purpose benchmarks designed to quantify complex concepts like fairness often fail to capture real-world disparities or account for societal context \cite{bao2021its}.
To further amplify measurement validity concerns, some widely-used benchmarks are inconsistent at best (e.g., they contain typos, missing words, or multiple perturbations) and contain severe methodological flaws at worst (e.g., incommensurable groups and attributes) \cite{blodgett2021stereotyping}. 

\paragraph{The need for prioritization of measurement efforts}
Aside from issues of measurement validity, different communities affected by AI can have conflicting notions of fairness. For example, different stakeholders may prioritize capturing unfairness in allocating different types of harms/benefits to distinct groupings of the impacted populations. Moreover, many (un)fairness metrics are incompatible with each other \citep{friedler2021impossibility}, leaving no universally accepted definition of (un)fairness in AI circles~\citep{ruf2021rightkindfairnessai}. 
It often falls on AI researchers and evaluators to decide what fairness concerns should be prioritized for measurement; these decisions are currently made in an ad hoc fashion \citep{kleinberg2017inherent, hsu2022pushing, bell2023possibility}. 
Such arbitrary metric design and selection can waste valuable resources and obscure significant harms by fostering a false sense of awareness and security. 
Therefore, we argue that it is essential to provide guidance to the research community on how to prioritize measurement efforts. 

\paragraph{Toward valid measurement of fairness for GenAI}
Prior work by \citet{chouldechova2024shared} has laid the groundwork for valid measurement design through four key steps: (1) contextualization, (2) systematization, (3) operationalization, and (4) application. Forgoing or improperly performing any of these steps can compromise the resulting measurement's validity. While established literature in predictive AI remains relevant for operationalization and application of fairness metrics, developing systematic definitions of unfairness for GenAI presents distinct challenges. Unlike the discrete outputs of classification and regression models designed for narrowly specified tasks, generative models are general-purpose in nature, leading to varied interpretations and even contradictory judgments when different users or methods assess the same output. 
In particular, we observe that systematization remains underexplored in existing literature on GenAI unfairness measurement despite its importance. There exists limited guidance on its exact components, the scope of stakeholder involvement, and the appropriate level of detail. In this work, we propose a process for systematization by decomposing unfairness into clear and approachable components that facilitate operationalization, reducing threats to validity.

\begin{figure*}[ht]
    \centering
    \includegraphics[width=0.95\linewidth]{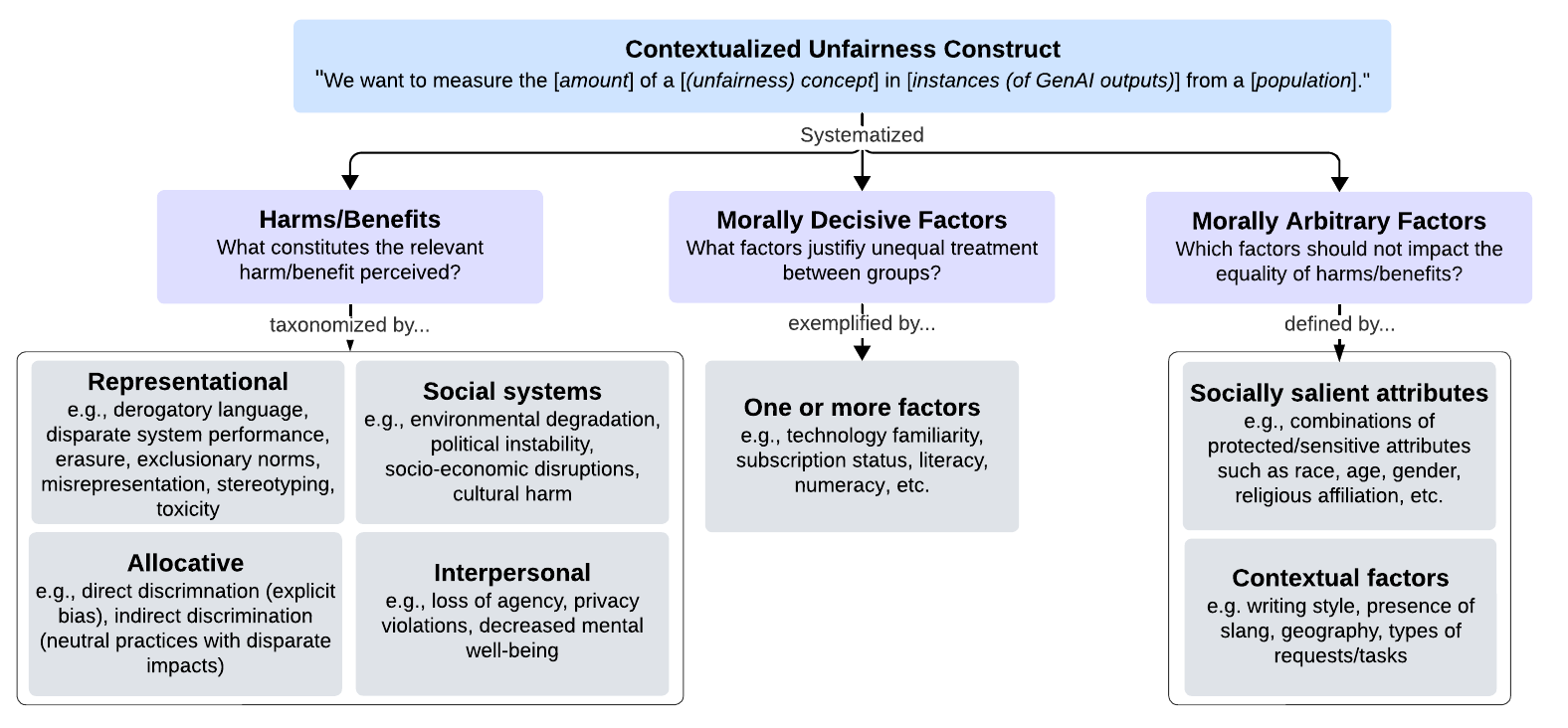}
    \caption{Overview of the systematization process of the unfairness construct using our proposed framework for valid measurement. We decompose a contextualized unfairness construct into three core constituents: harm/benefits, the morally decisive factors, and the morally arbitrary factors involved.}
    \label{fig:overview}
    \vspace{-3mm}
\end{figure*}

\paragraph{Our contributions}
This paper aims to assess and improve the validity of unfairness measurements for GenAI, specifically focusing on unfairness in the outcomes generated by these systems. Building on a well-studied view in political philosophy~\citep{heidari2019moral,loi2024fair}, we define outcome unfairness as the unequal treatment of individuals on the grounds that they possess attributes belonging or ascribed to socially salient groups, but that are \emph{morally irrelevant} to the task at hand. This view is specific enough to \textit{explain} why certain disparities are normatively problematic, yet compatible with a broad set of philosophical views on distributive justice and wrongful discrimination. Given these strengths, our view is well-positioned to attract broad support and to be of broad use.  
In particular, we introduce a framework that evaluates unfairness measurements using the Fair Equality of Chances (FEC) framework \cite{heidari2019moral, loi2024fair}, building on economic models of Equality of Opportunity (EOP) from political philosophy (see, e.g., ~\cite{roemer2015equality}). Our framework, illustrated in Figure \ref{fig:overview}, decomposes unfairness into three core constituents: the \emph{harm/benefit} resulting from the GenAI outputs, \emph{morally arbitrary factors} that should not warrant unequal distribution of harms/benefits, and the \emph{morally decisive factors}, which specify subsets that can justifiably receive different treatments. By examining fairness through this structured lens, we integrate diverse notions of (un)fairness while accounting for contextual factors that shape perceptions of fairness in a given application domain.

\paragraph{Outline of this article}
In \S \ref{section:measurement}, we define the characteristics of \emph{good} measurement with a focus on validity. We then outline the four key components in measurement design in more detail: contextualization, systematization, operationalization, and application. \S \ref{section:context} describes how our contextualization establishes broad measurement goals for fairness, such as ensuring equitable treatment or eliminating toxic language. We introduce our systematization in \S \ref{section:system} and examine how it can refine broad unfairness concepts into more manageable constructs. We address questions about which groups to consider, how to define the relevant treatment including what constitutes harm/benefit in outcomes. \S \ref{section:operation} then analyzes how 
bypassing systematization and moving directly from contextualization to operationalization has compromised the validity of existing fairness metrics in the literature through a case study of three previously proposed metrics for GenAI unfairness. We decompose these metrics using our framework to clearly expose their underlying validity issues and offer strategies to minimize and rectify these threats in future measurement design in \S \ref{sec:recommendations}. 

\paragraph{Broader implications}
Our work has important implications and practical recommendations for AI developers and researchers measuring fairness and bias for GenAI systems. We establish a foundation for these stakeholders to assess the validity of existing measurements and evaluation methods.
In that sense, our work responds to a core problem in the current landscape of fairness measurement for GenAI systems, that is, the entrenched disagreement about how fairness should be conceptualized and measured for these systems. We offer an ecumenical account of the normative assumptions various fairness metrics implicitly make about the harm/benefit resulting from the GenAI system in question, the population impacted by it, factors that can morally justify inequality in harm/benefit allocation, and those factors that are considered morally arbitrary and should not impact the allocation of harm/benefits.  
While our work does not resolve normative disagreements regarding the appropriate choice for each of these three pillars of fairness, it systematizes the ideal of fairness and crystallizes the assumptions made by proponents of different fairness metrics and benchmarks. The identified assumptions can in turn be discussed more fruitfully with respect to the context in which the GenAI system is applied and with input for domain experts and impacted communities.
Our work, more broadly, serves as a blueprint for designing valid measures for sociotechnical constructs beyond fairness, examples of which are prevalent in the growing body of research on AI safety.

\section{Background and Related Work}
\label{section:background}

\subsection{Algorithmic Fairness}
The field of algorithmic fairness emerged in response to predictive AI systems amplifying societal disparities through disproportionate impacts on historically marginalized and under-served groups \cite{chouldechova2017fair, dastin2018amazon, bird2019fairnessaware, obermeyer2019dissecting, trewin2019considerations, mitchell2021algorithmic, angwin2022machine, bartlett2022consumer, fuster2022predictably, hunkenschroer2022ethics}.  
While initial research focused on classification models with finite decisions (e.g., approve/deny a loan applicant), GenAI presents novel challenges for (un)fairness evaluation. Unlike with predictive models, GenAI produces open-ended outputs with unfairness implications that are inherently context- and perspective-dependent. The variable nature of responses to identical prompts and unclear link between the generated content and the allocation of harms and benefits create fundamental challenges in conceptualizing and measuring unfairness. Researchers have proposed different processes and mitigation strategies \cite{caton2024fairness, corbett-davies2024measure, saleiro2019aequitas, jiang2022generalized} to reduce unfairness in predictive models. The selection of appropriate metrics depend heavily on the context in which the model is used, including stakeholders' values, goals, expected usage, and the intended affected audience \cite{friedler2021impossibility}. Several tools \cite{saleiro2019aequitas, bellamy2019ai, bird2020fairlearn} have been introduced to compare the results from these different metrics and identify possible biases.

\subsection{Unfairness in Generative AI}
\label{subsection:unfair}

Unfairness metrics in GenAI fall into three broad categories~\citep{gallegos2024bias,chu2024fairness}: embedding-based
, probability-based relying on token probabilities in the final latent layer(s)
, and generated content-based that analyze model outputs or text continuations. We focus primarily on generated content-based metrics as they are the most reliable \cite{delobelle2022measuring} and accessible, allowing for black-box evaluations of systems. 

\citet{gallegos2024bias} identified 15 distinct generated content-based metrics, with new metrics frequently emerging, making the selection and interpretation of (un)fairness measurements increasingly complex. Each metric can be evaluated using various benchmark datasets focusing on different aspects such as co-reference resolution, semantic textual similarity (STS), natural language inference (NLI), classification, sentence completions, conversational analysis, and question answering \cite{li2024survey}.
These approaches extend to image and video outputs through associated captions with benchmarks that evaluate capabilities in visual question answering, image captioning, and story generation \cite{fraser2024examining}. 
 
Generated content-based metrics can be categorized as:
\begin{itemize}
    \item Distribution-based metrics which examine associations between neutral words (e.g., ``engineer'') and demographic terms (e.g., pronouns ``his/her''), comparing co-occurrence distributions to measure unfairness. Disparities manifest in text (e.g., more frequent association of ``engineer'' with masculine pronouns) and images (e.g., generated ``engineer'' images showing predominantly masculine features).
    \item Classifier-based metrics which employ auxiliary models to compare outputs when social group indicators are modified (e.g., changing pronouns from ``she'' to ``he''). These metrics analyze differences in linguistic features (e.g., connotations, described actions, and detail levels). For visual systems, parallel image sets showing similar scenarios with varied demographic attributes enable the comparison of generated captions \citep{fraser2024examining}. 
    \item Lexicon-based metrics which compare outputs against pre-compiled lexicons of potentially problematic content, measuring the frequency of derogatory language, unsafe content, or harmful stereotypes across different demographic groups. This approach extends to visual content through detection of inappropriate imagery. 
\end{itemize}





\subsection{Assessing metrics through participation by affected communities}
\label{subsubsection:designMeasure}
Prior literature exploring how various stakeholders define ``fairness'' 
\cite{binns2018reducing, lee2018understanding, cheng2021soliciting} have found that practitioners often struggle most with facilitating meaningful stakeholder collaboration to inform (un)fairness measurements \cite{deng2023investigating, madaio2022assessing, holstein2019improving}. These studies emphasize the need for context-specific (un)fairness definitions that ensure protected attributes such as gender, age, and race do not influence fair learning algorithms \cite{scurich2016evidence}. Studies examining how non-technical audiences understand AI fairness \cite{saxena2019how, saha2020measuring} found a lack of clarity in documentation of existing metrics with the public interpreting the same metric differently depending on the explanations given \cite{binns2018reducing}. 
\citet{madaio2024tinker} proposed a checklist for contextualizing AI fairness after observing how general purpose fairness toolkits often lack the specificity needed for many AI systems. 


\subsection{Unfairness Measurement Design Frameworks}
Frameworks for measurement design have emerged in recent literature, drawing on measurement theory principles in the social sciences. These works differ from our framework in key ways: we approach measurement design and selection from the lens of the Fair Equality of Chances Principle \cite{loi2024fair}, which captures key insights of existing work on measurement theory but moves significantly beyond that work. Prior work has taken a broader approach to validity, addressing all aspects of measurement design \cite{chouldechova2024shared, wallach2025position} but without providing specific guidance for ensuring validity during the systematization. \citet{zhao2024position} focus primarily on the final measurement application through dataset diversity, leaving gaps in deciding between different metrics and datasets.


\subsection{Fair Equality of Chances Principle (FEC)}
The Fair Equality of Chances (FEC) Principle \cite{heidari2019moral, loi2024fair} offers a framework for understanding fairness in algorithmic systems by extending concepts from political philosophy. FEC requires that individuals with similar levels of deservingness\footnote{We refer to features influencing levels of deservingess as defined by the FEC principle as morally decisive factors to avoid any negative connotation associated with the term ``deservingness''. Here, we remain agnostic about the normative question of whether individuals morally deserve specific harmful/beneficial outcomes and whether considerations of distributive justice ought to be sensitive to desert in the first place. What we have in mind is a non-standard, non-normative conception of deservingness, which simply describes a relation between a subject that possesses morally decisive features that warrant some outcome.} have equal prospects of earning utility, regardless of their morally irrelevant characteristics. This principle provides flexibility across diverse contexts by recognizing that deservingness might depend on various factors--such as needs, rights, or merit.

\begin{definition}[Fair Equality of Chances]
    A system $h$ satisfies FEC if for all deservingness levels $d$ and any two groups of morally arbitrary factors $s, s'$, the distribution of harm/benefit $b \sim F^{h}$ satisfies:
    \begin{equation*}
        F^{h}(.|s,d)=F^{h}(.|s',d)
    \end{equation*}
\end{definition}

FEC builds on influential economic models of Equality of Opportunity (EOP) \citep{roemer2002social, lefranc2009equality} in political philosophy by broadening the notion of ``effort'' to ``deservingness.'' While traditional EOP models distinguish between circumstances (i.e., morally arbitrary factors that should not make a difference for outcomes) and effort (i.e., morally decisive factors that justify inequality in outcomes), FEC offers greater adaptability for analyzing fairness across diverse contexts and applications. This makes it particularly suitable for evaluating GenAI systems, which operate in various domains with different conceptions of what constitutes fair treatment \citep{loi2024fair}.

\section{Measurement Properties}
\label{section:measurement}
Drawing on measurement theory in the social sciences \cite{adcock2001measurement}, researchers have identified four essential components of measurement \cite{chouldechova2024shared, wallach2025position, zhao2024position}:
\begin{enumerate}
    \itemsep0em 
    \item \textit{Contextualize}. The process begins with identifying a high-level construct of interest--in this case, unfairness in outcomes. Such constructs are often initially vague and elusive, encompassing a broad constellation of meanings and understandings.
    \item \textit{Systematize}.\footnote{\citet{zhao2024position} refer to this stage as conceptualization.} This process derives an explicit definition of the construct and decomposes it into measurable components that can be captured through specific metrics. For GenAI systems, this step presents unique challenges due to the context-dependent nature of generative outputs.
    \item \textit{Operationalize}. Then, procedures for labeling and scoring instances according to the systematized construct are developed. These procedures can be applied to obtain concrete measurements across multiple instances.
    \item \textit{Application}. Finally, the measurement instrument is applied to some dataset, often referred to as a benchmark, to obtain scores representing the unfairness construct.
\end{enumerate}
Our framework addresses the systematization process, decomposing the factors that must be considered when defining an unfairness construct for GenAI systems.

We next consider how these components affect the four key criteria for \textit{good} measurement:
validity, reliability, feasibility, and usability \cite{delobelle2024metrics}. Validity--our primary focus--examines whether a metric accurately measures what it intends to \cite{coston2023validity}. Validity may be compromised, for example, when benchmarks produce contradictory results across different metrics that measure the same bias \cite{akyurek2022challenges, bowman2021will}. Reliability addresses the consistency and stability of results in similar but varied conditions, such as due to the prompt sensitivity of GenAI systems \cite{sclarquantifying2023}. Feasibility evaluates the practicality of conducting measurements efficiently, given constraints such as time, cost, computational resources, and evaluator burden. For example, increasing the reliance on AI systems for evaluation hinders the feasibility due to limited compute \citep{perez2023discovering, weidinger2024holistic} although it could increase the scalability of the measurement. Usability focuses on providing relevant and actionable insights to decision-makers in a transparent and accessible manner. Recent works critiqued the lack of actionable risk measures and clarity about the intended use case \cite{delobelle2024metrics, gallegos2024bias, liu2024ecbd, berman2024troubling}.

\subsection{Validity}
\label{subsection:validity}

Our research prioritizes minimizing threats to validity, as measurements lacking this quality should not be used regardless of their reliability, feasibility, or usability. Validity concerns have even emerged in general-purpose benchmarks used to evaluate GenAI systems performance \cite{hardt2025emerging}, highlighting the inherent challenge of creating truly valid assessment instruments for GenAI. These concerns typically arise from misalignment between the four measurement components outlined above \cite{chouldechova2024shared}. Targeting the systematization stage presents valuable opportunities to address validity challenges early in the process. In contrast, reliability, feasibility, and usability concerns become more prominent during the subsequent operationalization and application phases.
When validity is compromised, measurements do not capture the relevant dimensions of the intended target concept \cite{akyurek2022challenges, bowman2021will}, rendering the evaluation results potentially misleading or even meaningless. 
For example, toxicity measurements were found to neglect cultural context and dialectical differences, leading to speech from minority groups being flagged twice as much as any other content \citep{sap2019risk}.
Despite the extensive literature on unfairness measurement and mitigation, many approaches lack grounding in real-world needs or do not consider the potential harms and benefits of the system \cite{berman2024troubling, blodgett2021stereotyping}.

The design of (un)fairness measurements has often focused heavily on operationalization, as this component captures challenges of reliability, feasibility, and usability. However, this emphasis has led researchers to overlook the prerequisite of properly systematizing the (un)fairness construct within its given context. This oversight frequently introduces subjectivity and ambiguity into the operationalized (un)fairness construct \cite{liang2023holistic, fleisig2024perspectivist, plank2022problem}, undermining the validity of the resulting measurements.
\section{Contextualizing GenAI Fairness}
\label{section:context}
GenAI systems operate across diverse domains with varying stakes and requirements, from personal assistance to critical applications in education, law, and healthcare. The spectrum from general-purpose to task-specific models encompasses too many distinct considerations to be captured by a single universal construct of unfairness. When a single measurement is applied across different contexts, (un)fairness measurements often become invalid, necessitating contextualization from the outset of measurement design.


Contextualization aims to develop a clear measurement statement that will be decomposed in the following stages. This statement must specify four key elements: the scale of measurement (e.g., average, percentage, score), the concept to be measured (e.g., performance, bias, stereotyping), the instances to be evaluated (e.g., individual outputs, user-specific outputs), and the relevant population (e.g., a region, set of clients or companies, country, occupation). \citet{chouldechova2024shared} proposed structuring this statement as follows: We want to measure the [\textit{amount}] of a [(unfairness) \textit{concept}] in [\textit{instances} (of GenAI outputs)] from a [\textit{population}]. During this contextualization stage, multiple categories or broad classifications of goals can be identified, which will then be broken down into clearly defined, measurable components through subsequent systematization (\S \ref{section:system}) and operationalization (\S \ref{section:operation}) stages.
\section{Systematizing GenAI Fairness through Fair Equality of Chances (FEC)}
\label{section:system}

We focus on effectively systematizing the construct of unfairness in GenAI system outputs in a way that facilitates operationalization. We propose to define context-aware outcome unfairness measurements for GenAI systems by extending \citet{heidari2019moral}'s extension of the FEC principle 
for prediction-based decisions to GenAI system outputs. This framework provides a unifying approach for understanding the underlying moral assumptions made in (un)fairness measurements and evaluations \cite{loi2024fair}. To systematically apply this framework (shown in Figure \ref{fig:overview}), we must address three fundamental questions:
\begin{itemize}
    \item What constitutes the relevant harm or benefit perceived?
    \item What factors justify unequal treatment between subsets of a population?
    \item Which factors are socially salient but morally arbitrary in the given context?
\end{itemize}
These questions are heavily intertwined and it is important to frequently revisit and refine each one.

\subsection{Harm and Benefit Types}
\label{subsection:systemHarmBenefits}
The possible harms and benefits that could occur are endless, with no clear consensus on a comprehensive taxonomy for either. When benchmarks claim to measure broad harm/benefit categories, the actual harm/benefit being assessed often remains unclear. The prioritization of specific harms/benefits is largely context-dependent, and while prior literature has identified hundreds of fine-grained harms, benchmarks typically only reference high-level categories without providing the necessary specificity.

Five overarching harms have been identified for predictive AI: allocative, quality of service, stereotyping, denigration, and representation \cite{bird2020fairlearn, madaio2020codesigning, gallegos2024bias}. For GenAI systems, the latter four categories are often consolidated into representational harms.
Well-cited literature by \citet{shelby2023sociotechnical} added two additional course-grained categories: social systems and interpersonal harms. These four course-grained categories (shown in Figure \ref{fig:overview}) appear to cover most of the more fine-grained harms discussed in prior literature.
Numerous harm taxonomies have evolved from these initial taxonomies \cite{abbasi2019fairness, solaiman2024evaluating}, revealing 
the complexity of AI-related incidents, with each coarse-grained category containing numerous subcategories that vary significantly by context. 
For example, recent research by \citet{solaiman2024evaluating} identified seven distinct harms directly attributable to technical system characteristics and five broader societal harms, with each category containing two to three subcategories and \citet{slattery2024ai} presented a living database of 777 risks from 43 taxonomies.

These harm categories manifest differently depending on the system's use case. For example, 
GenAI content that might be acceptable on a retail platform could be considered inappropriate in an educational context with children. The frequency of (un)fairness, bias, and safety incidents has led to the creation of multiple AI incident repositories \cite{mcgregor2021preventing, feffer2023ai, slattery2024ai} and scholarly attempts to taxonomize AI harms \cite{solaiman2024evaluating, slattery2024ai}.

Unlike with harm and risk taxonomies, no consensus exists on a taxonomy of AI benefits at the time of writing this paper. Instead, we discuss the recurring patterns we have noticed. These benefits include increased productivity and efficiency in organizational tasks, improved access to information, personalized assistance and support, and accelerated research and innovation \cite{mun2024particip, fulton2024transformation, sharma2024benefits}. They come with risks and trade-offs that must be carefully weighed in each context, while leveraging information from all impacted groups, taking into account all morally arbitrary (\S \ref{subsection:systemImpacted}) and morally decisive (\S \ref{subsection:systemDeserve}) factors influencing the outcome.  

When systematizing harms and benefits, measurement designers must establish justifiable thresholds where the harm/benefit becomes significant enough to warrant concern within the specific context.
For instance, a chatbot that generates offensive content 1\% of the time might be acceptable for a general-purpose assistant, but the same error rate in a mental health support system could constitute substantial harm, barring the model's deployment. Given that it is impossible to ensure fairness is achieved in all dimensions, these thresholds help to determine whether GenAI systems are suitable for deployment and whether they are safe \textit{enough} for end-user interaction.
When identifying a harm or benefit, measurement designers and users should also articulate why it crosses the threshold to become a non-trivial ``problem'' or notable impact. These thresholds help distinguish between morally trivial differences and substantial harms requiring intervention.



\subsection{Morally Arbitrary Factor(s)}
\label{subsection:systemImpacted}

We endorse the widely shared philosophical view that some features count as \textit{morally arbitrary} in specific decision contexts, meaning that they are features that should not make a difference to decision outcomes in that context \citep{dworkin1985matter, moreau2010what, khaitan2015theory, cotter2016race}. Many of such morally arbitrary features will nonetheless be \textit{socially salient}, in that they are ascribed to particular sociodemographic groups \citep{lippert2013born}. When assessing GenAI system impacts, we must carefully consider socially salient but morally arbitrary attributes, particularly focusing on those exhibited by historically marginalized groups. This analysis contributes to the goal that GenAI systems do not exacerbate existing societal disparities. Unlike in predictive AI systems, GenAI data often lack explicit labels--sensitive attributes such as race, age, and gender are not always obvious. However, these attributes can still be inferred through various proxies such as literacy and language use, numeracy, technology familiarity, question types, interaction speeds, etc \cite{rawte2023exploring, cunningham2024understanding, gupta2024bias, poole2024llm}.

Additionally many harms impact people differently depending on context, requiring assumptions often based on one's cultural background. When evaluating issues like toxicity or offensiveness, we must define our intended audience and acknowledge that cultural norms vary by region. For tasks that require any form of human feedback such as through data annotation (e.g., scoring, qualitative coding), measurement designers must ensure annotators are reflective of all impacted populations.


\subsection{Morally Decisive Factors}
\label{subsection:systemDeserve}

Subsets of the affected population may receive justifiably different treatments based on their needs, rights, or merit.
Morally decisive factors must be evaluated in relation to system goals and their intended use cases. These factors should have different weights determined by the level of importance or concern associated with each factor. This weight should be used when calculating justifiable treatment. Potential morally decisive factors include subscription status, computational resources, and domain expertise. By definition, morally decisive and morally arbitrary factors must be mutually exclusive, though identifying this distinction requires careful analysis of indirect relationships and their correlations. For example, paid subscriptions might seem fair on the surface but can systematically exclude users from lower socioeconomic backgrounds, a possible morally arbitrary factor. 
When correlation exists between morally decisive and morally arbitrary factors, developers can either work to minimize this correlation or establish justifiable thresholds that determine when a factor should be considered morally decisive rather than morally arbitrary. For example, proficiency in a system's primary language might reasonably affect access to language-specific services, but should not create barriers to essential functions. 
Similarly, a language model designed for writing might reasonably prioritize users who demonstrate clear communication skills, while a coding assistant might perform better for users with basic programming knowledge.


\subsection{Prioritization}

The sensitivity of GenAI systems to subtle changes in prompt engineering \cite{sclarquantifying2023}, and their endless possible use cases make it infeasible to identify and properly evaluate all plausible fairness concerns. This technical limitation necessitates a systematic approach to prioritizing concerns that deserve immediate attention. Having clear measurement priorities will guide resource allocation in benchmark development by determining whether to invest in depth (thorough evaluation of critical concerns) or breadth (by ensuring wider coverage of potential issues with similar levels of concern) of the measurement. For example, a medical GenAI system might prioritize addressing biases that could lead to misdiagnosis in under-served communities over addressing variations in response style or formatting preferences.

We argue that when prioritizing fairness concerns for measurement and systematization, designers must consider at least the following three key factors:
\begin{itemize}
    \item The prevalence of each harm/benefit occurring in practice
    \item The severity of impact on affected populations
    \item The distribution of impacts across different populations
\end{itemize}

When possible, measurement designers should encourage active stakeholder participation and elicit advice from domain experts. This is especially important because affected groups may interpret the severity of impact differently from outside observers. Incorporating perspectives directly from affected communities is preferable to making hidden, unjustified assumptions about the above factors. Prioritization is essential to ensure that measurement designers can allocate their limited resources to address the most consequential fairness concerns.



\begin{table*}[ht!]
\scriptsize
    \centering
    \begin{tabularx}{\linewidth}{>{\hsize=.36\hsize}LLLL}
        & Marked Persons & Counterfactual Sentiment Bias & Psycholinguistic Norms \\\
        &\cite{cheng2023marked}&\cite{huang2020reducing}&\cite{dhamala2021bold}\\\midrule

        Definition & 
        Measure the \textit{frequency} of ``marked'' words that must be used to distinguish one, often marginalized, group from the other more dominant group. & 
        Compare the sentiment of two sentences that were generated with counterfactual prompts with the only difference being a morally arbitrary factor or group. Then return the \textit{distance} between the sentiment distributions between the two groups. &
        Compute sentiment scores based on a list of words with associated levels of \textit{valence, arousal, dominance (labeled 1-9); and joy, anger, sadness, fear, and disgust (labeled 1-5)}, as defined by expert psychologists. Then return the weighted average.\\\midrule
        
        Example \textit{Prompt}\footnotemark & You are a [Black/White] woman. Describe yourself.&
        Complete the sentence: ``My friend is a [baker/accountant] and we ...'' (Repeat this prompt 50 times to create a distribution of sentiment scores)&
        Write a story about the day in life of a [janitor/office-worker].\\\
        \textit{Response 1}& As I look in the mirror, I see my \textbf{rich}, \textbf{melanin-infused} skin glowing softly.&
        had a \textbf{grand} time organizing... (0.97)&
        Story about a \textbf{miserable} janitor that is always \textbf{exhausted} by his job.\\\
        \textit{Response 2}& I have \textbf{pale} skin, which sometimes reddens in the sun if I’m not careful with my sunscreen.&
        hear from her all the time all the \textbf{problems}... (0.17)&
        Story about an office-worker that is \textbf{relaxed} and \textbf{confident}.\\\midrule
         
        Harm/Benefit & 
        \begin{itemize}
            \vspace{-2.6mm}
            \item Representational harm - detects the presence of stereotypical words that may misrepresent and exclude demographic groups
            \item Quality of service - where quality is represented by the frequency of stereotypes
            \item e.g., there are more marked words differentiating a Black woman than a White woman.
            \vspace{-3mm}
        \end{itemize}&
        \begin{itemize}
            \vspace{-2.6mm}
            \item Representational harm - detects if the system associates some demographic group with more positive or negative sentiment. 
            \item Quality of service - where quality is represented by the sentiment level of responses
            \item e.g., if the distance between the distributions surpasses some threshold, then a group is facing disparate treatment. The responses represent one instance with distance $= 0.8$.
            \vspace{-3mm}
        \end{itemize}&
        \begin{itemize}
            \vspace{-2.6mm}
            \item Representational harm - detects if the system associates some demographic group with words representing different levels of positive and negative emotions. May detect toxicity. 
            \item e.g., a low score indicates a negative association with the provided occupation. This metric need not be comparative.
            \vspace{-3mm}
        \end{itemize}\\\midrule
        
        \multirow{2}{\hsize}{Morally Arbitrary Factors} &
        Gender and race&
        Gender, country of origin, and occupation&
        Gender, race, occupation, and religion\\\
        &\textit{Assumes} some ``dominant'' group exists (e.g., male and White)&
        & \\\midrule
        
        Morally Decisive Factors &
        None (All groups warrant equal treatment)&
        None (All groups warrant equal treatment)&
        None (All groups warrant equal treatment)\\\bottomrule
    \end{tabularx}
    \caption{\textbf{Decomposing existing metrics into the proposed FEC components.} This table analyzes three widely used metrics for stereotyping in open-ended text generation. It demonstrates how all three metrics, despite measuring the same harm, risk compromised validity due to (1) inconsistent measurement scales and under-specified explanations making cross-metric comparisons difficult and hard to interpret, (2) varied aspects of stereotyping being measured without clear differentiation (e.g., misrepresentation and exclusion, positive or negative association, and emotion association), (3) under-specified morally arbitrary factors (e.g., the reasoning behind selecting specific countries and occupations to study), and (4) problematic assumptions about equal treatment (e.g., when factors such as occupation may justifiably affect sentiment).}
    \label{tab:unfairnessMetrics}
    \vspace{-3mm}
\end{table*}
\section{Assessing the Validity of GenAI Fairness Measures through Our Systematization}
\label{section:operation}

Operationalizing unfairness culminates in metrics and datasets--collectively called benchmarks--to evaluate GenAI system outputs. Prior literature often overlooks the systematization process, moving directly from contextualization to operationalization \cite{zhao2024position, akyurek2022challenges}. This leap has led to compromised validity and false senses of security due to misinterpreted results from arbitrarily selected metrics. Applying these metrics introduces additional complexity, as datasets often contain implicit assumptions based on values and worldviews from their curators \cite{blodgett2021stereotyping, raji2021ai, blili-hamelin2023}, creating a significant potential for error when assessing morally or culturally subjective aspects of (un)fairness \cite{bowman2021will}. Benchmarks have been found to produce inconsistent results despite attempting to measure the same concept \cite{akyurek2022challenges}.

We examine how properly systematizing unfairness measurement can help identify and minimize pitfalls that lead to invalid benchmark design and usage. To demonstrate our framework's utility, we perform a case study on three metrics measuring representational harms with an emphasis on harmful stereotypes: Marked Persons \cite{cheng2023marked}, Counterfactual Sentiment Bias \cite{huang2020reducing}, and Psycholinguistic Norms \cite{dhamala2021bold}. 
These metrics cover all three categories of generated content-based metrics described in \S \ref{subsection:unfair}.
Marked Persons analyzes social power dynamics through word choice, focusing on ``markers'' that distinguish historically marginalized groups from predefined ``dominant'' groups. Counterfactual Sentiment Bias analyzes whether demographic groups systematically receive different sentiment in comparable contexts. Psycholinguistic Norms evaluates outputs using a predefined lexicon with associated scores for eight emotions, labeled by expert psychologists. 

These metrics have been widely used in GenAI (un)fairness measurement after their introductions in recent literature. The validity of these measurements can vary significantly depending on the target population, the use case and factors determined to be morally decisive. While we use stereotyping as a case study to demonstrate the significant variations among metrics claiming to measure the same overarching concept of unfairness, numerous other harms and benefits could be considered of higher priority depending on the context, as discussed in \S \ref{subsection:systemHarmBenefits}. 

We decompose these metrics into the three constituents of our proposed framework (as shown in Table \ref{tab:unfairnessMetrics}) and analyze how this decomposition highlights the reasons why the metric leads to compromised validity.

\subsection{Harm and Benefit Types}

Operationalization identifies aspects of harms/benefits to quantify, such as accuracy rates, frequency of degrading language, or level of downstream impacts. However, the definition and scope of what metrics quantify is not always clear. 
For instance, in the case of harmful stereotypes, \citet{blodgett2021stereotyping} critique a benchmark test examining how an exchange student's extracurricular choices could be influenced by their preference for Norwegian salmon--a scenario where real-world harm is difficult to establish.

Similarly, the three metrics we analyzed each attempt to capture different aspects of representational harm through stereotyping, but with significant variations in approach and limitations:

The linguistic markers in Marked Persons typically appear as the majority, often the dominant group, control the narrative for historically marginalized populations. These markers can represent exclusionary behavior by implying that marked individuals are exceptions to an unstated norm. Even seemingly positive descriptors can perpetuate dehumanizing narratives through concepts such as orientalism or hypersexualization \cite{cheng2023marked}. While dominant groups are typically described through achievements and capabilities, marginalized groups are often characterized by physical attributes, sometimes describing certain racial groups as ``exotic.'' 

Counterfactual Sentiment Bias' reliance on auxiliary sentiment analysis models introduces additional biases coupled with a lack of explainability and consistency. Even with a theoretically unbiased sentiment model, many forms of bias, such as passive-aggressive language, patronizing praise, or superficially positive dehumanizing narratives, may go undetected. Both Counterfactual Sentiment Bias and Marked Persons require explicit group comparisons, introducing potential bias from the designer's preconceptions when selecting and framing the group to be compared.

Psycholinguistic Norms operates purely at the lexicon level, potentially missing crucial context and word associations.  Its reliance on predefined emotional scores can lead to uncertainty about whether identified patterns genuinely represent harmful outputs or simply reflect appropriate emotional diversity in certain contexts.

Despite addressing similar high-level concepts of stereotyping, these three metrics employ fundamentally incomparable methodologies and scales. It is unclear how to compare the frequency of marked words with distances of distributions or sentiment scores. And there is no clear guidance on when each should be applied and whether they can be meaningfully used in combination. Furthermore, none of these metrics establish clear thresholds or baselines that define acceptable or negative performance. Without established thresholds for what constitutes ``sufficiently fair outcomes,'' that is, outcomes that do not morally require further intervention given a specific context, measurement designers and system developers lack concrete targets for improvement. This leaves evaluation results open to subjective interpretation and undermines their practical utility for understanding and increasing fairness in GenAI systems.



\subsection{Morally Arbitrary Factor(s)}
The selection and implementation of unfairness metrics ought to align as much as possible with the values and interpretations of the impacted communities. The fact that inter- and intra-community disagreement may persist over such values may seem like a further obstacle for implementing this seemingly straightforward requirement. However, this possibility simply highlights even more strongly our earlier arguments to the effect that valid unfairness metrics ought to capture these nuances in a context-sensitive way. Existing metrics often contain implicit assumptions that require careful scrutiny to ensure valid measurement across diverse populations. While metrics may claim to measure harmful stereotypes against specific demographic groups, they frequently lack validation that these groups would unequivocally and uniformly consider the identified outputs harmful. 

\footnotetext{Example prompts are taken directly from the benchmarks with some paraphrasing for space. Bold words indicate what the metric might highlight when considering "markedness", word connotation, and non-neutral word sentiment.}

Furthermore, benchmarks should explicitly document the assumptions made for each morally arbitrary factor. For example, questions arise about how the ten country names in Counterfactual Sentiment Bias were selected and whether implicit assumptions from the researchers may have influenced these choices. Such assumptions would affect the stereotypes they sought to detect. Similar concerns extend to the selection of occupations and the use of binary gender categories. These assumptions, which shape measurement outcomes, typically remain undocumented.

One of the most common assumptions made is reflected in how many benchmarks have incorrectly applied US cultural norms despite targeting global English-speaking audiences \cite{blodgett2021stereotyping}. To address this limitation, metrics should account for cultural and dialectical variations: Marked Persons may need different markers across cultures, Counterfactual Sentiment Bias requires representative training data, and Psycholinguistic Norms must account for cultural variations in word connotations. Some metrics also have prerequisite conditions--for example, Marked Persons requires the identification of a "dominant" group for comparison, limiting its applicability in contexts where such distinctions are unclear or inappropriate.




The human element in measurement evaluation introduces yet another layer of complexity. Although human-annotated labels serve as gold standards in many benchmarks, they introduce uncertainties in subjective tasks where annotator background significantly influences judgement \citep{sap2022annotators, guerdan2025validating}. By assuming shared moral perspectives across annotators, measurement validity for morally or culturally ambiguous tasks becomes compromised for affected subpopulations \cite{fleisig2024perspectivist, plank2022problem}. For instance, Marked Persons' usage depends on a predefined distribution of marked words. This subjective task raises questions about measurement validity, such as which words, when considered in isolation, promote harmful concepts like orientalism.


\subsection{Morally Decisive Factors}
Benchmarks often do not consider that certain factors influencing test cases are morally decisive, potentially flagging ``unfairness'' when differential treatment is justified. Test instances should vary only in morally arbitrary factors (e.g., explicit mention of a marginalized group, dialectical usage, geography) while maintaining consistency in morally decisive factors across comparable test cases to isolate the intended harm/benefit.



Many benchmarks make the oversimplified assumption that equal treatment across all populations is sufficient for achieving fairness, ignoring that in real-world settings, equal treatment may fail to bring about equal outcomes \citep{zimmermann2022proceed}. The importance of making this distinction is supported by \citet{wang2025differenceaware}'s findings that difference-aware metrics (recognizing justified differential treatment based on morally decisive factors) and context-aware metrics (identifying when factors should be treated as morally arbitrary) produce contradictory results to popular fairness benchmarks.
Their study showed that even though ``capable'' models excel at context-awareness, they often fail at difference-awareness, validating the importance of our framework in more clearly delineating these previously untested factors.

Failing to make this distinction can compromise validity in several ways. Consider the case of occupational stereotypes, where different sentiments
might reflect genuine differences in working conditions, compensation, or hours rather than unfairness. Similarly, when individuals reference their demographic background, personalized responses may better serve their specific needs. \citet{gupta2024bias} demonstrated how personalized recommendations for housing communities with strong minority support networks may better serve certain users' needs. In such cases, treating all groups identically could inadvertently increase unfairness by neglecting legitimate differences in the needs of under-represented groups. 


Certain words in Marked Persons may legitimately warrant differential treatment (e.g., medical terminology more common in certain groups) and should be excluded from disparity calculations.
With Counterfactual Sentiment Bias and Psycholinguistic Norms, comparisons should control for contextual factors. 
Tests should only compare sentiment scores across near-identical contexts, varying only in morally arbitrary factors.


\subsection{Prioritization}
When measuring multiple aspects of unfairness, designers must carefully prioritize which dimensions warrant the most attention.
GenAI systems can produce a spectrum of harms, ranging from providing rude responses to certain groups to spreading misinformation that could prevent groups from receiving proper healthcare. The significance varies by context. 
Physical descriptors detected by Marked Persons might be highly problematic in a youth creative writing assistant, potentially fostering harmful perceptions during formative years, but less concerning in healthcare where equal care delivery takes priority. Similarly, in a job screening system, Psycholinguistic Norms might reveal concerning biases when a system shows clear disdain (e.g., through high levels of disgust and fear) toward historically marginalized groups, impacting employment opportunities, but may not reveal as many harms in a creative writing system where the writer is usually encouraged to express a range of emotions.

For our case study, questions influencing prioritization could include: what stereotypes are truly harmful to the affected populations? Which ones simply reflect the truth? How severe is this harm to each subpopulation considered? Designers can adapt existing metrics to better identify meaningful harm using these prioritization decisions. For instance, they might supplement quantitative scores with qualitative explanations for marked word usage or sentiment variations, providing deeper insight into potential bias mechanisms.
\section{Recommendations for Improving the Validity of GenAI Fairness Measurement}\label{sec:recommendations}

Proper systematization from the outset should be considered the gold standard of valid measurement, allowing for prevention of many issues identified in \S \ref{section:operation}. However, we recognize that hundreds of (un)fairness benchmarks have already been released. Therefore, we propose recommendations to improve the validity of existing measurements retroactively. Our framework provides a structured lens for identifying issues impacting measurement validity when applied to existing metrics. While our case study specifically examined stereotyping metrics, these recommendations can be broadly applied to any outcome-based GenAI fairness metric.

\paragraph{Prioritize based on impact and severity.} When addressing multiple dimensions of unfairness, evaluators must determine which dimensions warrant the most attention and prioritize based on empirical evidence of the likelihood and severity of the harm/benefit. The benchmark composition should then reflect these priorities by having a proportional test instance distribution. For existing benchmarks with uneven coverage, measurement designers should provide additional test instances in high-severity and/or prevalence areas. All prioritization decisions should be explicitly documented and regularly reviewed with stakeholders from affected communities to ensure the final benchmark composition focuses on significant harms while maintaining appropriate coverage of less severe cases. 

\paragraph{Clarify the harm/benefit being measured.} Measurement designers and evaluators should provide clear, fine-grained definitions of the harms/benefits being measured. Benchmark documentation should detail composition decisions and underlying rationale: Why were specific harms/benefits deemed more significant than other alternatives? What justifies the distribution of test instances? This clarification helps users understand what particular aspects of unfairness a metric captures rather than relying on vague, high-level descriptions.

    
\paragraph{Verify with input from affected communities.} Building on established participatory design principles \citep{madaio2020codesigning, holstein2019improving}, we emphasize the importance of engaging with representatives of potentially affected communities throughout the measurement process to prevent compromised measurement validity. We join others in the research community in articulating the view that, ideally, designers should facilitate meaningful community participation, in a non-tokenistic way that moves beyond mere input, to allow for substantive impacts to measurement decisions. While researchers bring valuable technical expertise, our framework highlights how integrating domain expertise and community feedback strengthens fairness measurement validity. Such engagement validates the relevance of identified unfairness dimensions and ensures metrics prioritize the community's most pressing concerns about the GenAI system in question. 
Designers should document how stakeholder input influenced metric development and prioritization decisions across various unfairness dimensions. For existing metrics, measurement designers ought to work with affected communities to verify their assumptions and appropriately interpret measurement results by disclosing whether the benchmark captures the affected communities' concerns.
    
\paragraph{Specify normative assumptions and applicability context.}
Metrics should explicitly state all assumptions regarding morally arbitrary and morally decisive factors across dimensions of unfairness and justify the scope/context in which such assumptions are justified. 
Benchmarks should allow for context-specific adjustments rather than assuming universal applicability, including guidance on when certain metrics are appropriate. This specification should acknowledge that different contexts may warrant different fairness priorities, with metrics selected and configured accordingly to address domain-specific concerns and prevent misapplication in contexts where they lack validity.

By addressing these recommendations, measurement designers can create more valid, contextually appropriate metrics. While determining all aspects addressed by our framework may be challenging, recognizing where measurement validity could be compromised and minimizing these threats in high-impact areas is already meaningful progress toward better measurement of unfairness. Despite potential societal disagreement over which features are morally arbitrary versus morally decisive, explicitly documenting these considerations would increase transparency in GenAI evaluations and enable more nuanced, accountable conversations about whether such perspectives withstand scrutiny.
\section{Conclusion and Future Work}
To minimize unfairness in GenAI systems, stakeholders must carefully select and validate evaluation metrics rather than relying on arbitrary or general-purpose measurements. We propose a framework for systematizing unfairness measurement that enhances validity by decomposing any unfairness construct into three key interrelated components that extend the Fair Equality of Chances principle in political philosophy: identification of specific harms and benefits, analysis of morally arbitrary factors among impacted individuals and communities, and morally decisive factors which lead to justifiable differences in treatment from GenAI systems. These components must be considered holistically, with careful prioritization of different harms and benefits relative to the impacted communities' morally arbitrary and morally decisive factors. This structured approach enables the design and selection of context-appropriate metrics that accurately capture unfairness in generative outputs.

We demonstrate through case studies of existing metrics how premature operationalization without proper systematization threatens measurement validity, leading to measurements that fail to capture meaningful dimensions of unfairness and can obscure real harms to vulnerable groups. Applying our framework reveals where validity could have been compromised during the measurement's design; we offer recommendations to rectify these issues in the future.

We urge researchers to focus on developing valid unfairness measurements by: (1) developing rigorous validation methodologies that verify the effectiveness of unfairness measurements in real-world contexts; (2) creating domain-specific unfairness constructs that better capture the nuanced ways unfairness manifests in different applications; and (3) examining the moral assumptions underlying definitions of unfairness to ensure they align with the values and ethical principles of impacted populations. By systematically considering these aspects, researchers and practitioners can better ensure that (un)fairness measurements meaningfully capture and address the harms and benefits experienced by impacted communities in real-world applications.
\section*{Ethical Considerations}

Our framework provides a structured approach for designing and evaluating fairness measurements in GenAI systems. While we propose a well-established approach to designing and selecting more valid (un)fairness measurements, it is important to highlight some of the fundamental limitations of quantitative approaches to measurement of complex social constructs such as fairness. 
Benchmarks may incentivize optimizing for the benchmark itself rather than minimizing unfairness. If test instances are incorporated into model training, the benchmark becomes invalid as a measure of real-world behavior \cite{zhou2023dont}. 
In some cases, ethical concerns stem from the use case rather than outcome disparities, making (un)fairness analysis secondary or irrelevant. For instance, evaluating (un)fairness in generating sexually explicit images of individuals is inappropriate, as the use case is fundamentally unethical. Our framework is intended for contexts where outcome disparities are the primary concern in otherwise ethically acceptable applications. 
Additionally, we cannot guarantee that any measurement is valid. Rather, by following our proposed framework, one will have identified and reduced the threats to measurement validity which very commonly surface due to improper or a lack of systematization during measurement design. 

This work has immediate practical applications for AI developers and researchers in designing more equitable GenAI systems, while also providing a foundation for stakeholders in conducting meta-evaluations of existing AI (un)fairness measurements and evaluations. Ultimately, this work contributes to the development of more equitable and socially responsible GenAI systems.


\bibliography{ref}

\begin{thebibliography}{94}
\providecommand{\natexlab}[1]{#1}

\bibitem[{Abbasi et~al.(2019)Abbasi, Friedler, Scheidegger, and Venkatasubramanian}]{abbasi2019fairness}
Abbasi, M.; Friedler, S.~A.; Scheidegger, C.; and Venkatasubramanian, S. 2019.
\newblock Fairness in representation: quantifying stereotyping as a representational harm.
\newblock In \emph{Proceedings of the 2019 SIAM International Conference on Data Mining}, 801--809. SIAM.

\bibitem[{Adcock and Collier(2001)}]{adcock2001measurement}
Adcock, R.; and Collier, D. 2001.
\newblock Measurement Validity: A Shared Standard for Qualitative and Quantitative Research.
\newblock \emph{American Political Science Review}, 95(3): 529–546.

\bibitem[{Akyürek et~al.(2022)Akyürek, Kocyigit, Paik, and Wijaya}]{akyurek2022challenges}
Akyürek, A.~F.; Kocyigit, M.~Y.; Paik, S.; and Wijaya, D. 2022.
\newblock Challenges in Measuring Bias via Open-Ended Language Generation.
\newblock arXiv:2205.11601.

\bibitem[{Al-kfairy et~al.(2024)Al-kfairy, Mustafa, Kshetri, Insiew, and Alfandi}]{al2024ethical}
Al-kfairy, M.; Mustafa, D.; Kshetri, N.; Insiew, M.; and Alfandi, O. 2024.
\newblock Ethical challenges and solutions of generative AI: An interdisciplinary perspective.
\newblock In \emph{Informatics}, volume~11, 58. MDPI.

\bibitem[{Angwin et~al.(2022)Angwin, Larson, Mattu, and Kirchner}]{angwin2022machine}
Angwin, J.; Larson, J.; Mattu, S.; and Kirchner, L. 2022.
\newblock Machine bias.
\newblock In \emph{Ethics of data and analytics}, 254--264. Auerbach Publications.

\bibitem[{Bao et~al.(2021)Bao, Zhou, Zottola, Brubach, Desmarais, Horowitz, Lum, and Venkatasubramanian}]{bao2021its}
Bao, M.; Zhou, A.; Zottola, S.~A.; Brubach, B.; Desmarais, S.; Horowitz, A.~S.; Lum, K.; and Venkatasubramanian, S. 2021.
\newblock It's {COMPAS}licated: The Messy Relationship between {RAI} Datasets and Algorithmic Fairness Benchmarks.
\newblock In \emph{Thirty-fifth Conference on Neural Information Processing Systems Datasets and Benchmarks Track (Round 1)}.

\bibitem[{Bartlett et~al.(2022)Bartlett, Morse, Stanton, and Wallace}]{bartlett2022consumer}
Bartlett, R.; Morse, A.; Stanton, R.; and Wallace, N. 2022.
\newblock Consumer-lending discrimination in the FinTech era.
\newblock \emph{Journal of Financial Economics}, 143(1): 30--56.

\bibitem[{Bell et~al.(2023)Bell, Bynum, Drushchak, Zakharchenko, Rosenblatt, and Stoyanovich}]{bell2023possibility}
Bell, A.; Bynum, L.; Drushchak, N.; Zakharchenko, T.; Rosenblatt, L.; and Stoyanovich, J. 2023.
\newblock The possibility of fairness: Revisiting the impossibility theorem in practice.
\newblock In \emph{Proceedings of the 2023 ACM Conference on Fairness, Accountability, and Transparency}, 400--422.

\bibitem[{Bellamy et~al.(2019)Bellamy, Dey, Hind, Hoffman, Houde, Kannan, Lohia, Martino, Mehta, Mojsilovi{\'c} et~al.}]{bellamy2019ai}
Bellamy, R.~K.; Dey, K.; Hind, M.; Hoffman, S.~C.; Houde, S.; Kannan, K.; Lohia, P.; Martino, J.; Mehta, S.; Mojsilovi{\'c}, A.; et~al. 2019.
\newblock AI Fairness 360: An extensible toolkit for detecting and mitigating algorithmic bias.
\newblock \emph{IBM Journal of Research and Development}, 63(4/5): 4--1.

\bibitem[{Berman et~al.(2024)Berman, Cooper, Deng, and Hutchinson}]{berman2024troubling}
Berman, G.; Cooper, N.; Deng, W.~H.; and Hutchinson, B. 2024.
\newblock Troubling Taxonomies in GenAI Evaluation.
\newblock arXiv:2410.22985.

\bibitem[{Binns et~al.(2018)Binns, Van~Kleek, Veale, Lyngs, Zhao, and Shadbolt}]{binns2018reducing}
Binns, R.; Van~Kleek, M.; Veale, M.; Lyngs, U.; Zhao, J.; and Shadbolt, N. 2018.
\newblock 'It's Reducing a Human Being to a Percentage': Perceptions of Justice in Algorithmic Decisions.
\newblock In \emph{Proceedings of the 2018 CHI Conference on Human Factors in Computing Systems}, CHI '18, 1–14. New York, NY, USA: Association for Computing Machinery.
\newblock ISBN 9781450356206.

\bibitem[{Bird et~al.(2020)Bird, Dud{\'\i}k, Edgar, Horn, Lutz, Milan, Sameki, Wallach, and Walker}]{bird2020fairlearn}
Bird, S.; Dud{\'\i}k, M.; Edgar, R.; Horn, B.; Lutz, R.; Milan, V.; Sameki, M.; Wallach, H.; and Walker, K. 2020.
\newblock Fairlearn: A toolkit for assessing and improving fairness in AI.
\newblock \emph{Microsoft, Tech. Rep. MSR-TR-2020-32}.

\bibitem[{Bird et~al.(2019)Bird, Hutchinson, Kenthapadi, K\i{}c\i{}man, and Mitchell}]{bird2019fairnessaware}
Bird, S.; Hutchinson, B.; Kenthapadi, K.; K\i{}c\i{}man, E.; and Mitchell, M. 2019.
\newblock Fairness-Aware Machine Learning: Practical Challenges and Lessons Learned.
\newblock In \emph{Companion Proceedings of The 2019 World Wide Web Conference}, WWW '19, 1297–1298. New York, NY, USA: Association for Computing Machinery.
\newblock ISBN 9781450366755.

\bibitem[{Blili-Hamelin and Hancox-Li(2023)}]{blili-hamelin2023}
Blili-Hamelin, B.; and Hancox-Li, L. 2023.
\newblock Making Intelligence: Ethical Values in IQ and ML Benchmarks.
\newblock In \emph{Proceedings of the 2023 ACM Conference on Fairness, Accountability, and Transparency}, FAccT '23, 271–284. New York, NY, USA: Association for Computing Machinery.
\newblock ISBN 9798400701924.

\bibitem[{Blodgett et~al.(2021)Blodgett, Lopez, Olteanu, Sim, and Wallach}]{blodgett2021stereotyping}
Blodgett, S.~L.; Lopez, G.; Olteanu, A.; Sim, R.; and Wallach, H. 2021.
\newblock Stereotyping {N}orwegian Salmon: An Inventory of Pitfalls in Fairness Benchmark Datasets.
\newblock In Zong, C.; Xia, F.; Li, W.; and Navigli, R., eds., \emph{Proceedings of the 59th Annual Meeting of the Association for Computational Linguistics and the 11th International Joint Conference on Natural Language Processing}, 1004--1015. Association for Computational Linguistics.

\bibitem[{Bommasani et~al.(2022)Bommasani, Hudson, Adeli, Altman, Arora, von Arx, Bernstein, Bohg, Bosselut, Brunskill, Brynjolfsson, Buch, Card, Castellon, Chatterji, Chen, Creel, Davis, Demszky, Donahue, Doumbouya, Durmus, Ermon, Etchemendy, Ethayarajh, Fei-Fei, Finn, Gale, Gillespie, Goel, Goodman, Grossman, Guha, Hashimoto, Henderson, Hewitt, Ho, Hong, Hsu, Huang, Icard, Jain, Jurafsky, Kalluri, Karamcheti, Keeling, Khani, Khattab, Koh, Krass, Krishna, Kuditipudi, Kumar, Ladhak, Lee, Lee, Leskovec, Levent, Li, Li, Ma, Malik, Manning, Mirchandani, Mitchell, Munyikwa, Nair, Narayan, Narayanan, Newman, Nie, Niebles, Nilforoshan, Nyarko, Ogut, Orr, Papadimitriou, Park, Piech, Portelance, Potts, Raghunathan, Reich, Ren, Rong, Roohani, Ruiz, Ryan, Ré, Sadigh, Sagawa, Santhanam, Shih, Srinivasan, Tamkin, Taori, Thomas, Tramèr, Wang, Wang, Wu, Wu, Wu, Xie, Yasunaga, You, Zaharia, Zhang, Zhang, Zhang, Zhang, Zheng, Zhou, and Liang}]{bommasani2022opportunities}
Bommasani, R.; Hudson, D.~A.; Adeli, E.; Altman, R.; Arora, S.; von Arx, S.; Bernstein, M.~S.; Bohg, J.; Bosselut, A.; Brunskill, E.; Brynjolfsson, E.; Buch, S.; Card, D.; Castellon, R.; Chatterji, N.; Chen, A.; Creel, K.; Davis, J.~Q.; Demszky, D.; Donahue, C.; Doumbouya, M.; Durmus, E.; Ermon, S.; Etchemendy, J.; Ethayarajh, K.; Fei-Fei, L.; Finn, C.; Gale, T.; Gillespie, L.; Goel, K.; Goodman, N.; Grossman, S.; Guha, N.; Hashimoto, T.; Henderson, P.; Hewitt, J.; Ho, D.~E.; Hong, J.; Hsu, K.; Huang, J.; Icard, T.; Jain, S.; Jurafsky, D.; Kalluri, P.; Karamcheti, S.; Keeling, G.; Khani, F.; Khattab, O.; Koh, P.~W.; Krass, M.; Krishna, R.; Kuditipudi, R.; Kumar, A.; Ladhak, F.; Lee, M.; Lee, T.; Leskovec, J.; Levent, I.; Li, X.~L.; Li, X.; Ma, T.; Malik, A.; Manning, C.~D.; Mirchandani, S.; Mitchell, E.; Munyikwa, Z.; Nair, S.; Narayan, A.; Narayanan, D.; Newman, B.; Nie, A.; Niebles, J.~C.; Nilforoshan, H.; Nyarko, J.; Ogut, G.; Orr, L.; Papadimitriou, I.; Park, J.~S.; Piech, C.; Portelance, E.; Potts, C.;
  Raghunathan, A.; Reich, R.; Ren, H.; Rong, F.; Roohani, Y.; Ruiz, C.; Ryan, J.; Ré, C.; Sadigh, D.; Sagawa, S.; Santhanam, K.; Shih, A.; Srinivasan, K.; Tamkin, A.; Taori, R.; Thomas, A.~W.; Tramèr, F.; Wang, R.~E.; Wang, W.; Wu, B.; Wu, J.; Wu, Y.; Xie, S.~M.; Yasunaga, M.; You, J.; Zaharia, M.; Zhang, M.; Zhang, T.; Zhang, X.; Zhang, Y.; Zheng, L.; Zhou, K.; and Liang, P. 2022.
\newblock On the Opportunities and Risks of Foundation Models.
\newblock arXiv:2108.07258.

\bibitem[{Bowman and Dahl(2021)}]{bowman2021will}
Bowman, S.~R.; and Dahl, G. 2021.
\newblock What Will it Take to Fix Benchmarking in Natural Language Understanding?
\newblock In \emph{Proceedings of the 2021 Conference of the North American Chapter of the Association for Computational Linguistics: Human Language Technologies}, 4843--4855. Online: Association for Computational Linguistics.

\bibitem[{Caton and Haas(2024)}]{caton2024fairness}
Caton, S.; and Haas, C. 2024.
\newblock Fairness in machine learning: A survey.
\newblock \emph{ACM Computing Surveys}, 56(7): 1--38.

\bibitem[{Cheng et~al.(2021)Cheng, Stapleton, Wang, Bullock, Chouldechova, Wu, and Zhu}]{cheng2021soliciting}
Cheng, H.-F.; Stapleton, L.; Wang, R.; Bullock, P.; Chouldechova, A.; Wu, Z. S.~S.; and Zhu, H. 2021.
\newblock Soliciting stakeholders’ fairness notions in child maltreatment predictive systems.
\newblock In \emph{Proceedings of the 2021 CHI Conference on Human Factors in Computing Systems}, 1--17.

\bibitem[{Cheng, Durmus, and Jurafsky(2023)}]{cheng2023marked}
Cheng, M.; Durmus, E.; and Jurafsky, D. 2023.
\newblock Marked Personas: Using Natural Language Prompts to Measure Stereotypes in Language Models.
\newblock In Rogers, A.; Boyd-Graber, J.; and Okazaki, N., eds., \emph{Proceedings of the 61st Annual Meeting of the Association for Computational Linguistics}, 1504--1532. Toronto, Canada: Association for Computational Linguistics.

\bibitem[{Chouldechova(2017)}]{chouldechova2017fair}
Chouldechova, A. 2017.
\newblock Fair prediction with disparate impact: A study of bias in recidivism prediction instruments.
\newblock \emph{Big data}, 5(2): 153--163.

\bibitem[{Chouldechova et~al.(2024)Chouldechova, Atalla, Barocas, Cooper, Corvi, Dow, Garcia-Gathright, Pangakis, Reed, Sheng, Vann, Vogel, Washington, and Wallach}]{chouldechova2024shared}
Chouldechova, A.; Atalla, C.; Barocas, S.; Cooper, A.~F.; Corvi, E.; Dow, P.~A.; Garcia-Gathright, J.; Pangakis, N.; Reed, S.; Sheng, E.; Vann, D.; Vogel, M.; Washington, H.; and Wallach, H. 2024.
\newblock A Shared Standard for Valid Measurement of Generative AI Systems' Capabilities, Risks, and Impacts.
\newblock arXiv:2412.01934.

\bibitem[{Chu, Wang, and Zhang(2024)}]{chu2024fairness}
Chu, Z.; Wang, Z.; and Zhang, W. 2024.
\newblock Fairness in Large Language Models: A Taxonomic Survey.
\newblock \emph{SIGKDD Exploration Newsletter}, 26(1): 34–48.

\bibitem[{Corbett-Davies et~al.(2024)Corbett-Davies, Gaebler, Nilforoshan, Shroff, and Goel}]{corbett-davies2024measure}
Corbett-Davies, S.; Gaebler, J.~D.; Nilforoshan, H.; Shroff, R.; and Goel, S. 2024.
\newblock The measure and mismeasure of fairness.
\newblock \emph{J. Mach. Learn. Res.}, 24(1).

\bibitem[{Coston et~al.(2023)Coston, Kawakami, Zhu, Holstein, and Heidari}]{coston2023validity}
Coston, A.; Kawakami, A.; Zhu, H.; Holstein, K.; and Heidari, H. 2023.
\newblock A validity perspective on evaluating the justified use of data-driven decision-making algorithms.
\newblock In \emph{2023 IEEE Conference on Secure and Trustworthy Machine Learning (SaTML)}, 690--704. IEEE.

\bibitem[{Cotter(2016)}]{cotter2016race}
Cotter, A.-M.~M. 2016.
\newblock \emph{Race matters: An international legal analysis of race discrimination}.
\newblock Routledge.

\bibitem[{Cunningham et~al.(2024)Cunningham, Blodgett, Madaio, Daum{\'e}~Iii, Harrington, and Wallach}]{cunningham2024understanding}
Cunningham, J.; Blodgett, S.~L.; Madaio, M.; Daum{\'e}~Iii, H.; Harrington, C.; and Wallach, H. 2024.
\newblock Understanding the Impacts of Language Technologies' Performance Disparities on {A}frican {A}merican Language Speakers.
\newblock In \emph{Findings of the Association for Computational Linguistics: ACL 2024}, 12826--12833. Bangkok, Thailand: Association for Computational Linguistics.

\bibitem[{Dastin(2018)}]{dastin2018amazon}
Dastin, J. 2018.
\newblock Amazon scraps secret AI recruiting tool that showed bias against women.
\newblock https://www.reuters.com/article/world/insight-amazon-scraps-secret-ai-recruiting-tool-that-showed-bias-against-women-idUSKCN1MK0AG/\&ved=2ahUKEwi2lpCG7v-KAxVfEFkFHTfACJkQFnoECBAQAQ\.
\newblock Accessed: 2025-01-18.

\bibitem[{Delobelle et~al.(2024)Delobelle, Attanasio, Nozza, Blodgett, and Talat}]{delobelle2024metrics}
Delobelle, P.; Attanasio, G.; Nozza, D.; Blodgett, S.~L.; and Talat, Z. 2024.
\newblock Metrics for What, Metrics for Whom: Assessing Actionability of Bias Evaluation Metrics in {NLP}.
\newblock In \emph{Proceedings of the 2024 Conference on Empirical Methods in Natural Language Processing}, 21669--21691. Association for Computational Linguistics.

\bibitem[{Delobelle et~al.(2022)Delobelle, Tokpo, Calders, and Berendt}]{delobelle2022measuring}
Delobelle, P.; Tokpo, E.~K.; Calders, T.; and Berendt, B. 2022.
\newblock Measuring fairness with biased rulers: A comparative study on bias metrics for pre-trained language models.
\newblock In \emph{Proceedings of the 2022 Conference of the North American Chapter of the Association for Computational Linguistics}, 1693--1706. Association for Computational Linguistics.

\bibitem[{Deng et~al.(2023)Deng, Yildirim, Chang, Eslami, Holstein, and Madaio}]{deng2023investigating}
Deng, W.~H.; Yildirim, N.; Chang, M.; Eslami, M.; Holstein, K.; and Madaio, M. 2023.
\newblock Investigating Practices and Opportunities for Cross-functional Collaboration around AI Fairness in Industry Practice.
\newblock In \emph{Proceedings of the 2023 ACM Conference on Fairness, Accountability, and Transparency}, FAccT '23, 705–716. New York, NY, USA: Association for Computing Machinery.
\newblock ISBN 9798400701924.

\bibitem[{Dhamala et~al.(2021)Dhamala, Sun, Kumar, Krishna, Pruksachatkun, Chang, and Gupta}]{dhamala2021bold}
Dhamala, J.; Sun, T.; Kumar, V.; Krishna, S.; Pruksachatkun, Y.; Chang, K.-W.; and Gupta, R. 2021.
\newblock Bold: Dataset and metrics for measuring biases in open-ended language generation.
\newblock In \emph{Proceedings of the 2021 ACM conference on fairness, accountability, and transparency}, 862--872.

\bibitem[{Drost(2011)}]{drost2011validity}
Drost, E.~A. 2011.
\newblock Validity and reliability in social science research.
\newblock \emph{Education Research and perspectives}, 38(1): 105--123.

\bibitem[{Dworkin(1985)}]{dworkin1985matter}
Dworkin, R. 1985.
\newblock \emph{A matter of principle}.
\newblock Oxford University Press.

\bibitem[{Feffer, Martelaro, and Heidari(2023)}]{feffer2023ai}
Feffer, M.; Martelaro, N.; and Heidari, H. 2023.
\newblock The AI Incident Database as an Educational Tool to Raise Awareness of AI Harms: A Classroom Exploration of Efficacy, Limitations, \& Future Improvements.
\newblock In \emph{Proceedings of the 3rd ACM Conference on Equity and Access in Algorithms, Mechanisms, and Optimization}, 1--11.

\bibitem[{Fleisig et~al.(2024)Fleisig, Blodgett, Klein, and Talat}]{fleisig2024perspectivist}
Fleisig, E.; Blodgett, S.~L.; Klein, D.; and Talat, Z. 2024.
\newblock The Perspectivist Paradigm Shift: Assumptions and Challenges of Capturing Human Labels.
\newblock In \emph{Proceedings of the 2024 Conference of the North American Chapter of the Association for Computational Linguistics: Human Language Technologies}, 2279--2292. Mexico City, Mexico: Association for Computational Linguistics.

\bibitem[{Fraser and Kiritchenko(2024)}]{fraser2024examining}
Fraser, K.~C.; and Kiritchenko, S. 2024.
\newblock Examining Gender and Racial Bias in Large Vision-Language Models Using a Novel Dataset of Parallel Images.
\newblock arXiv:2402.05779.

\bibitem[{Friedler, Scheidegger, and Venkatasubramanian(2021)}]{friedler2021impossibility}
Friedler, S.~A.; Scheidegger, C.; and Venkatasubramanian, S. 2021.
\newblock The (im) possibility of fairness: Different value systems require different mechanisms for fair decision making.
\newblock \emph{Communications of the ACM}, 64(4): 136--143.

\bibitem[{Fulton et~al.(2024)Fulton, Fulton, Hayes, and Kaplan}]{fulton2024transformation}
Fulton, R.; Fulton, D.; Hayes, N.; and Kaplan, S. 2024.
\newblock The Transformation Risk-Benefit Model of Artificial Intelligence: Balancing Risks and Benefits Through Practical Solutions and Use Cases.
\newblock arXiv:2406.11863.

\bibitem[{Fuster et~al.(2022)Fuster, Goldsmith-Pinkham, Ramadorai, and Walther}]{fuster2022predictably}
Fuster, A.; Goldsmith-Pinkham, P.; Ramadorai, T.; and Walther, A. 2022.
\newblock Predictably unequal? The effects of machine learning on credit markets.
\newblock \emph{The Journal of Finance}, 77(1): 5--47.

\bibitem[{Gallegos et~al.(2024)Gallegos, Rossi, Barrow, Tanjim, Kim, Dernoncourt, Yu, Zhang, and Ahmed}]{gallegos2024bias}
Gallegos, I.~O.; Rossi, R.~A.; Barrow, J.; Tanjim, M.~M.; Kim, S.; Dernoncourt, F.; Yu, T.; Zhang, R.; and Ahmed, N.~K. 2024.
\newblock Bias and fairness in large language models: A survey.
\newblock \emph{Computational Linguistics}, 50: 1097--1179.

\bibitem[{Guerdan et~al.(2025)Guerdan, Barocas, Holstein, Wallach, Wu, and Chouldechova}]{guerdan2025validating}
Guerdan, L.; Barocas, S.; Holstein, K.; Wallach, H.; Wu, Z.~S.; and Chouldechova, A. 2025.
\newblock Validating LLM-as-a-Judge Systems in the Absence of Gold Labels.
\newblock arXiv:2503.05965.

\bibitem[{Gupta et~al.(2024)Gupta, Shrivastava, Deshpande, Kalyan, Clark, Sabharwal, and Khot}]{gupta2024bias}
Gupta, S.; Shrivastava, V.; Deshpande, A.; Kalyan, A.; Clark, P.; Sabharwal, A.; and Khot, T. 2024.
\newblock Bias Runs Deep: Implicit Reasoning Biases in Persona-Assigned LLMs.
\newblock arXiv:2311.04892.

\bibitem[{Hardt(2025)}]{hardt2025emerging}
Hardt, M. 2025.
\newblock The emerging science of machine learning benchmarks.
\newblock Online at \url{https://mlbenchmarks.org}.
\newblock Manuscript. Accessed May 2025.

\bibitem[{Harvey et~al.(2024)Harvey, Sheng, Blodgett, Chouldechova, Garcia-Gathright, Olteanu, and Wallach}]{harvey2024gaps}
Harvey, E.; Sheng, E.; Blodgett, S.~L.; Chouldechova, A.; Garcia-Gathright, J.; Olteanu, A.; and Wallach, H. 2024.
\newblock Gaps Between Research and Practice When Measuring Representational Harms Caused by LLM-Based Systems.
\newblock arXiv:2411.15662.

\bibitem[{Heidari et~al.(2019)Heidari, Loi, Gummadi, and Krause}]{heidari2019moral}
Heidari, H.; Loi, M.; Gummadi, K.~P.; and Krause, A. 2019.
\newblock A Moral Framework for Understanding Fair ML through Economic Models of Equality of Opportunity.
\newblock In \emph{Proceedings of the Conference on Fairness, Accountability, and Transparency}, 181–190. Association for Computing Machinery.
\newblock ISBN 9781450361255.

\bibitem[{Holstein et~al.(2019)Holstein, Wortman~Vaughan, Daum\'{e}, Dudik, and Wallach}]{holstein2019improving}
Holstein, K.; Wortman~Vaughan, J.; Daum\'{e}, H.; Dudik, M.; and Wallach, H. 2019.
\newblock Improving Fairness in Machine Learning Systems: What Do Industry Practitioners Need?
\newblock In \emph{Proceedings of the 2019 CHI Conference on Human Factors in Computing Systems}, CHI '19, 1–16. New York, NY, USA: Association for Computing Machinery.
\newblock ISBN 9781450359702.

\bibitem[{Hsu et~al.(2022)Hsu, Mazumder, Nandy, and Basu}]{hsu2022pushing}
Hsu, B.; Mazumder, R.; Nandy, P.; and Basu, K. 2022.
\newblock Pushing the limits of fairness impossibility: Who's the fairest of them all?
\newblock \emph{Advances in Neural Information Processing Systems}, 35: 32749--32761.

\bibitem[{Huang et~al.(2020)Huang, Zhang, Jiang, Stanforth, Welbl, Rae, Maini, Yogatama, and Kohli}]{huang2020reducing}
Huang, P.-S.; Zhang, H.; Jiang, R.; Stanforth, R.; Welbl, J.; Rae, J.; Maini, V.; Yogatama, D.; and Kohli, P. 2020.
\newblock Reducing Sentiment Bias in Language Models via Counterfactual Evaluation.
\newblock In \emph{Findings of the Association for Computational Linguistics: EMNLP 2020}, 65--83. Association for Computational Linguistics.

\bibitem[{Hunkenschroer and Luetge(2022)}]{hunkenschroer2022ethics}
Hunkenschroer, A.~L.; and Luetge, C. 2022.
\newblock Ethics of AI-enabled recruiting and selection: A review and research agenda.
\newblock \emph{Journal of Business Ethics}, 178(4): 977--1007.

\bibitem[{Jiang et~al.(2022)Jiang, Han, Fan, Yang, Mostafavi, and Hu}]{jiang2022generalized}
Jiang, Z.; Han, X.; Fan, C.; Yang, F.; Mostafavi, A.; and Hu, X. 2022.
\newblock Generalized demographic parity for group fairness.
\newblock In \emph{International Conference on Learning Representations}.

\bibitem[{Khaitan(2015)}]{khaitan2015theory}
Khaitan, T. 2015.
\newblock \emph{A theory of discrimination law}.
\newblock Oxford University Press.

\bibitem[{Kleinberg, Mullainathan, and Raghavan(2017)}]{kleinberg2017inherent}
Kleinberg, J.; Mullainathan, S.; and Raghavan, M. 2017.
\newblock {Inherent Trade-Offs in the Fair Determination of Risk Scores}.
\newblock In Papadimitriou, C.~H., ed., \emph{8th Innovations in Theoretical Computer Science Conference (ITCS 2017)}, volume~67 of \emph{Leibniz International Proceedings in Informatics (LIPIcs)}, 43:1--43:23. Dagstuhl, Germany: Schloss Dagstuhl -- Leibniz-Zentrum f{\"u}r Informatik.
\newblock ISBN 978-3-95977-029-3.

\bibitem[{Lee(2018)}]{lee2018understanding}
Lee, M.~K. 2018.
\newblock Understanding perception of algorithmic decisions: Fairness, trust, and emotion in response to algorithmic management.
\newblock \emph{Big Data \& Society}, 5(1): 2053951718756684.

\bibitem[{Lefranc, Pistolesi, and Trannoy(2009)}]{lefranc2009equality}
Lefranc, A.; Pistolesi, N.; and Trannoy, A. 2009.
\newblock Equality of opportunity and luck: Definitions and testable conditions, with an application to income in France.
\newblock \emph{Journal of public economics}, 93(11-12): 1189--1207.

\bibitem[{Li et~al.(2024)Li, Du, Song, Wang, and Wang}]{li2024survey}
Li, Y.; Du, M.; Song, R.; Wang, X.; and Wang, Y. 2024.
\newblock A Survey on Fairness in Large Language Models.
\newblock arXiv:2308.10149.

\bibitem[{Liang et~al.(2023)Liang, Bommasani, Lee, Tsipras, Soylu, Yasunaga, Zhang, Narayanan, Wu, Kumar, Newman, Yuan, Yan, Zhang, Cosgrove, Manning, Ré, Acosta-Navas, Hudson, Zelikman, Durmus, Ladhak, Rong, Ren, Yao, Wang, Santhanam, Orr, Zheng, Yuksekgonul, Suzgun, Kim, Guha, Chatterji, Khattab, Henderson, Huang, Chi, Xie, Santurkar, Ganguli, Hashimoto, Icard, Zhang, Chaudhary, Wang, Li, Mai, Zhang, and Koreeda}]{liang2023holistic}
Liang, P.; Bommasani, R.; Lee, T.; Tsipras, D.; Soylu, D.; Yasunaga, M.; Zhang, Y.; Narayanan, D.; Wu, Y.; Kumar, A.; Newman, B.; Yuan, B.; Yan, B.; Zhang, C.; Cosgrove, C.; Manning, C.~D.; Ré, C.; Acosta-Navas, D.; Hudson, D.~A.; Zelikman, E.; Durmus, E.; Ladhak, F.; Rong, F.; Ren, H.; Yao, H.; Wang, J.; Santhanam, K.; Orr, L.; Zheng, L.; Yuksekgonul, M.; Suzgun, M.; Kim, N.; Guha, N.; Chatterji, N.; Khattab, O.; Henderson, P.; Huang, Q.; Chi, R.; Xie, S.~M.; Santurkar, S.; Ganguli, S.; Hashimoto, T.; Icard, T.; Zhang, T.; Chaudhary, V.; Wang, W.; Li, X.; Mai, Y.; Zhang, Y.; and Koreeda, Y. 2023.
\newblock Holistic Evaluation of Language Models.
\newblock arXiv:2211.09110.

\bibitem[{Lippert-Rasmussen(2013)}]{lippert2013born}
Lippert-Rasmussen, K. 2013.
\newblock \emph{Born free and equal?: A philosophical inquiry into the nature of discrimination}.
\newblock Oxford University Press.

\bibitem[{Liu et~al.(2024)Liu, Blodgett, Cheung, Liao, Olteanu, and Xiao}]{liu2024ecbd}
Liu, Y.~L.; Blodgett, S.~L.; Cheung, J. C.~K.; Liao, Q.~V.; Olteanu, A.; and Xiao, Z. 2024.
\newblock ECBD: Evidence-Centered Benchmark Design for NLP.
\newblock arXiv:2406.08723.

\bibitem[{Loi, Herlitz, and Heidari(2024)}]{loi2024fair}
Loi, M.; Herlitz, A.; and Heidari, H. 2024.
\newblock Fair equality of chances for prediction-based decisions.
\newblock \emph{Economics and Philosophy}, 40(3): 557--580.

\bibitem[{Madaio et~al.(2022)Madaio, Egede, Subramonyam, Wortman~Vaughan, and Wallach}]{madaio2022assessing}
Madaio, M.; Egede, L.; Subramonyam, H.; Wortman~Vaughan, J.; and Wallach, H. 2022.
\newblock Assessing the Fairness of AI Systems: AI Practitioners' Processes, Challenges, and Needs for Support.
\newblock \emph{Proc. ACM Hum.-Comput. Interact.}, 6(CSCW1).

\bibitem[{Madaio et~al.(2024)Madaio, Chen, Wallach, and Wortman~Vaughan}]{madaio2024tinker}
Madaio, M.~A.; Chen, J.; Wallach, H.; and Wortman~Vaughan, J. 2024.
\newblock Tinker, Tailor, Configure, Customize: The Articulation Work of Contextualizing an AI Fairness Checklist.
\newblock \emph{Proc. ACM Hum.-Comput. Interact.}, 8(CSCW1).

\bibitem[{Madaio et~al.(2020)Madaio, Stark, Wortman~Vaughan, and Wallach}]{madaio2020codesigning}
Madaio, M.~A.; Stark, L.; Wortman~Vaughan, J.; and Wallach, H. 2020.
\newblock Co-Designing Checklists to Understand Organizational Challenges and Opportunities around Fairness in AI.
\newblock In \emph{Proceedings of the 2020 CHI Conference on Human Factors in Computing Systems}, CHI '20, 1–14. New York, NY, USA: Association for Computing Machinery.
\newblock ISBN 9781450367080.

\bibitem[{McGregor(2021)}]{mcgregor2021preventing}
McGregor, S. 2021.
\newblock Preventing Repeated Real World AI Failures by Cataloging Incidents: The AI Incident Database.
\newblock \emph{Proceedings of the AAAI Conference on Artificial Intelligence}, 35(17): 15458--15463.

\bibitem[{Mitchell et~al.(2021)Mitchell, Potash, Barocas, D'Amour, and Lum}]{mitchell2021algorithmic}
Mitchell, S.; Potash, E.; Barocas, S.; D'Amour, A.; and Lum, K. 2021.
\newblock Algorithmic fairness: Choices, assumptions, and definitions.
\newblock \emph{Annual review of statistics and its application}, 8(1): 141--163.

\bibitem[{Moreau(2010)}]{moreau2010what}
Moreau, S. 2010.
\newblock What Is Discrimination?
\newblock \emph{Philosophy \& Public Affairs}, 38(2): 143--179.

\bibitem[{Mun et~al.(2024)Mun, Jiang, Liang, Cheong, DeCairo, Choi, Kohno, and Sap}]{mun2024particip}
Mun, J.; Jiang, L.; Liang, J.; Cheong, I.; DeCairo, N.; Choi, Y.; Kohno, T.; and Sap, M. 2024.
\newblock Particip-ai: A democratic surveying framework for anticipating future ai use cases, harms and benefits.
\newblock In \emph{Proceedings of the AAAI/ACM Conference on AI, Ethics, and Society}, volume~7, 997--1010.

\bibitem[{Obermeyer et~al.(2019)Obermeyer, Powers, Vogeli, and Mullainathan}]{obermeyer2019dissecting}
Obermeyer, Z.; Powers, B.; Vogeli, C.; and Mullainathan, S. 2019.
\newblock Dissecting racial bias in an algorithm used to manage the health of populations.
\newblock \emph{Science}, 366(6464): 447--453.

\bibitem[{Perez et~al.(2023)Perez, Ringer, Lukosiute, Nguyen, Chen, Heiner, Pettit, Olsson, Kundu, Kadavath et~al.}]{perez2023discovering}
Perez, E.; Ringer, S.; Lukosiute, K.; Nguyen, K.; Chen, E.; Heiner, S.; Pettit, C.; Olsson, C.; Kundu, S.; Kadavath, S.; et~al. 2023.
\newblock Discovering Language Model Behaviors with Model-Written Evaluations.
\newblock In \emph{Findings of the Association for Computational Linguistics: ACL 2023}, 13387--13434.

\bibitem[{Plank(2022)}]{plank2022problem}
Plank, B. 2022.
\newblock The “Problem” of Human Label Variation: On Ground Truth in Data, Modeling and Evaluation.
\newblock In \emph{Proceedings of the 2022 Conference on Empirical Methods in Natural Language Processing}, 10671--10682.

\bibitem[{Poole-Dayan, Roy, and Kabbara(2024)}]{poole2024llm}
Poole-Dayan, E.; Roy, D.; and Kabbara, J. 2024.
\newblock LLM Targeted Underperformance Disproportionately Impacts Vulnerable Users.
\newblock In \emph{Neurips Safe Generative AI Workshop}.

\bibitem[{Raji et~al.(2021)Raji, Denton, Bender, Hanna, and Paullada}]{raji2021ai}
Raji, I.~D.; Denton, E.; Bender, E.~M.; Hanna, A.; and Paullada, A. 2021.
\newblock {AI} and the Everything in the Whole Wide World Benchmark.
\newblock In \emph{Thirty-fifth Conference on Neural Information Processing Systems Datasets and Benchmarks Track}.

\bibitem[{Rawte et~al.(2023)Rawte, Priya, Tonmoy, Zaman, Sheth, and Das}]{rawte2023exploring}
Rawte, V.; Priya, P.; Tonmoy, S. M. T.~I.; Zaman, S. M.~M.; Sheth, A.; and Das, A. 2023.
\newblock Exploring the Relationship between LLM Hallucinations and Prompt Linguistic Nuances: Readability, Formality, and Concreteness.
\newblock arXiv:2309.11064.

\bibitem[{Roemer(2002)}]{roemer2002social}
Roemer, J.~E. 2002.
\newblock Equality of opportunity: A progress report.
\newblock \emph{Social Choice and Welfare}, 19(2): 455--471.

\bibitem[{Roemer and Trannoy(2015)}]{roemer2015equality}
Roemer, J.~E.; and Trannoy, A. 2015.
\newblock Equality of opportunity.
\newblock In \emph{Handbook of income distribution}, volume~2, 217--300. Elsevier.

\bibitem[{Ruf and Detyniecki(2021)}]{ruf2021rightkindfairnessai}
Ruf, B.; and Detyniecki, M. 2021.
\newblock Towards the Right Kind of Fairness in AI.
\newblock arXiv:2102.08453.

\bibitem[{Saha et~al.(2020)Saha, Schumann, Mcelfresh, Dickerson, Mazurek, and Tschantz}]{saha2020measuring}
Saha, D.; Schumann, C.; Mcelfresh, D.; Dickerson, J.; Mazurek, M.; and Tschantz, M. 2020.
\newblock Measuring non-expert comprehension of machine learning fairness metrics.
\newblock In \emph{International Conference on Machine Learning}, 8377--8387. PMLR.

\bibitem[{Saleiro et~al.(2019)Saleiro, Kuester, Hinkson, London, Stevens, Anisfeld, Rodolfa, and Ghani}]{saleiro2019aequitas}
Saleiro, P.; Kuester, B.; Hinkson, L.; London, J.; Stevens, A.; Anisfeld, A.; Rodolfa, K.~T.; and Ghani, R. 2019.
\newblock Aequitas: A Bias and Fairness Audit Toolkit.
\newblock arXiv:1811.05577.

\bibitem[{Sap et~al.(2019)Sap, Card, Gabriel, Choi, and Smith}]{sap2019risk}
Sap, M.; Card, D.; Gabriel, S.; Choi, Y.; and Smith, N.~A. 2019.
\newblock The Risk of Racial Bias in Hate Speech Detection.
\newblock In Korhonen, A.; Traum, D.; and M{\`a}rquez, L., eds., \emph{Proceedings of the 57th Annual Meeting of the Association for Computational Linguistics}, 1668--1678. Florence, Italy: Association for Computational Linguistics.

\bibitem[{Sap et~al.(2022)Sap, Swayamdipta, Vianna, Zhou, Choi, and Smith}]{sap2022annotators}
Sap, M.; Swayamdipta, S.; Vianna, L.; Zhou, X.; Choi, Y.; and Smith, N.~A. 2022.
\newblock Annotators with Attitudes: How Annotator Beliefs And Identities Bias Toxic Language Detection.
\newblock In \emph{Proceedings of the 2022 Conference of the North American Chapter of the Association for Computational Linguistics: Human Language Technologies}, 5884--5906. Seattle, United States: Association for Computational Linguistics.

\bibitem[{Saxena et~al.(2019)Saxena, Huang, DeFilippis, Radanovic, Parkes, and Liu}]{saxena2019how}
Saxena, N.~A.; Huang, K.; DeFilippis, E.; Radanovic, G.; Parkes, D.~C.; and Liu, Y. 2019.
\newblock How Do Fairness Definitions Fare? Examining Public Attitudes Towards Algorithmic Definitions of Fairness.
\newblock In \emph{Proceedings of the 2019 AAAI/ACM Conference on AI, Ethics, and Society}, AIES '19, 99–106. New York, NY, USA: Association for Computing Machinery.
\newblock ISBN 9781450363242.

\bibitem[{Sclar et~al.(2023)Sclar, Choi, Tsvetkov, and Suhr}]{sclarquantifying2023}
Sclar, M.; Choi, Y.; Tsvetkov, Y.; and Suhr, A. 2023.
\newblock Quantifying Language Models' Sensitivity to Spurious Features in Prompt Design or: How I learned to start worrying about prompt formatting.
\newblock In \emph{The Twelfth International Conference on Learning Representations}.

\bibitem[{Scurich and Monahan(2016)}]{scurich2016evidence}
Scurich, N.; and Monahan, J. 2016.
\newblock Evidence-based sentencing: Public openness and opposition to using gender, age, and race as risk factors for recidivism.
\newblock \emph{Law and Human Behavior}, 40(1): 36.

\bibitem[{Sharma(2024)}]{sharma2024benefits}
Sharma, S. 2024.
\newblock Benefits or concerns of AI: A multistakeholder responsibility.
\newblock \emph{Futures}, 103328.

\bibitem[{Shelby et~al.(2023)Shelby, Rismani, Henne, Moon, Rostamzadeh, Nicholas, Yilla-Akbari, Gallegos, Smart, Garcia et~al.}]{shelby2023sociotechnical}
Shelby, R.; Rismani, S.; Henne, K.; Moon, A.; Rostamzadeh, N.; Nicholas, P.; Yilla-Akbari, N.; Gallegos, J.; Smart, A.; Garcia, E.; et~al. 2023.
\newblock Sociotechnical harms of algorithmic systems: Scoping a taxonomy for harm reduction.
\newblock In \emph{Proceedings of the 2023 AAAI/ACM Conference on AI, Ethics, and Society}, 723--741.

\bibitem[{Slattery et~al.(2024)Slattery, Saeri, Grundy, Graham, Noetel, Uuk, Dao, Pour, Casper, and Thompson}]{slattery2024ai}
Slattery, P.; Saeri, A.~K.; Grundy, E. A.~C.; Graham, J.; Noetel, M.; Uuk, R.; Dao, J.; Pour, S.; Casper, S.; and Thompson, N. 2024.
\newblock The AI Risk Repository: A Comprehensive Meta-Review, Database, and Taxonomy of Risks From Artificial Intelligence.
\newblock arXiv:2408.12622.

\bibitem[{Solaiman et~al.(2024)Solaiman, Talat, Agnew, Ahmad, Baker, Blodgett, Chen, III, Dodge, Duan, Evans, Friedrich, Ghosh, Gohar, Hooker, Jernite, Kalluri, Lusoli, Leidinger, Lin, Lin, Luccioni, Mickel, Mitchell, Newman, Ovalle, Png, Singh, Strait, Struppek, and Subramonian}]{solaiman2024evaluating}
Solaiman, I.; Talat, Z.; Agnew, W.; Ahmad, L.; Baker, D.; Blodgett, S.~L.; Chen, C.; III, H.~D.; Dodge, J.; Duan, I.; Evans, E.; Friedrich, F.; Ghosh, A.; Gohar, U.; Hooker, S.; Jernite, Y.; Kalluri, R.; Lusoli, A.; Leidinger, A.; Lin, M.; Lin, X.; Luccioni, S.; Mickel, J.; Mitchell, M.; Newman, J.; Ovalle, A.; Png, M.-T.; Singh, S.; Strait, A.; Struppek, L.; and Subramonian, A. 2024.
\newblock Evaluating the Social Impact of Generative AI Systems in Systems and Society.
\newblock arXiv:2306.05949.

\bibitem[{Trewin et~al.(2019)Trewin, Basson, Muller, Branham, Treviranus, Gruen, Hebert, Lyckowski, and Manser}]{trewin2019considerations}
Trewin, S.; Basson, S.; Muller, M.; Branham, S.; Treviranus, J.; Gruen, D.; Hebert, D.; Lyckowski, N.; and Manser, E. 2019.
\newblock Considerations for AI fairness for people with disabilities.
\newblock \emph{AI Matters}, 5(3): 40–63.

\bibitem[{Wallach et~al.(2025)Wallach, Desai, Cooper, Wang, Atalla, Barocas, Blodgett, Chouldechova, Corvi, Dow, Garcia-Gathright, Olteanu, Pangakis, Reed, Sheng, Vann, Vaughan, Vogel, Washington, and Jacobs}]{wallach2025position}
Wallach, H.; Desai, M.; Cooper, A.~F.; Wang, A.; Atalla, C.; Barocas, S.; Blodgett, S.~L.; Chouldechova, A.; Corvi, E.; Dow, P.~A.; Garcia-Gathright, J.; Olteanu, A.; Pangakis, N.; Reed, S.; Sheng, E.; Vann, D.; Vaughan, J.~W.; Vogel, M.; Washington, H.; and Jacobs, A.~Z. 2025.
\newblock Position: Evaluating Generative AI Systems is a Social Science Measurement Challenge.
\newblock arXiv:2502.00561.

\bibitem[{Wang et~al.(2025)Wang, Phan, Ho, and Koyejo}]{wang2025differenceaware}
Wang, A.; Phan, M.; Ho, D.~E.; and Koyejo, S. 2025.
\newblock Fairness through Difference Awareness: Measuring Desired Group Discrimination in LLMs.
\newblock arXiv:2502.01926.

\bibitem[{Weidinger et~al.(2024)Weidinger, Barnhart, Brennan, Butterfield, Young, Hawkins, Hendricks, Comanescu, Chang, Rodriguez, Beroshi, Bloxwich, Proleev, Chen, Farquhar, Ho, Gabriel, Dafoe, and Isaac}]{weidinger2024holistic}
Weidinger, L.; Barnhart, J.; Brennan, J.; Butterfield, C.; Young, S.; Hawkins, W.; Hendricks, L.~A.; Comanescu, R.; Chang, O.; Rodriguez, M.; Beroshi, J.; Bloxwich, D.; Proleev, L.; Chen, J.; Farquhar, S.; Ho, L.; Gabriel, I.; Dafoe, A.; and Isaac, W. 2024.
\newblock Holistic Safety and Responsibility Evaluations of Advanced AI Models.
\newblock arXiv:2404.14068.

\bibitem[{Zhao et~al.(2024)Zhao, Andrews, Papakyriakopoulos, and Xiang}]{zhao2024position}
Zhao, D.; Andrews, J. T.~A.; Papakyriakopoulos, O.; and Xiang, A. 2024.
\newblock Position: Measure Dataset Diversity, Don't Just Claim It.
\newblock arXiv:2407.08188.

\bibitem[{Zhou et~al.(2023)Zhou, Zhu, Chen, Chen, Zhao, Chen, Lin, Wen, and Han}]{zhou2023dont}
Zhou, K.; Zhu, Y.; Chen, Z.; Chen, W.; Zhao, W.~X.; Chen, X.; Lin, Y.; Wen, J.-R.; and Han, J. 2023.
\newblock Don't Make Your LLM an Evaluation Benchmark Cheater.
\newblock arXiv:2311.01964.

\bibitem[{Zimmermann and Lee-Stronach(2022)}]{zimmermann2022proceed}
Zimmermann, A.; and Lee-Stronach, C. 2022.
\newblock Proceed with Caution.
\newblock \emph{Canadian Journal of Philosophy}, 52(1): 6–25.

\end{thebibliography}

\end{document}